\documentclass[11pt]{article}

\usepackage{acl}

\usepackage{times}
\usepackage{latexsym}
\usepackage[T1]{fontenc}
\usepackage[utf8]{inputenc}
\usepackage{microtype}
\usepackage{inconsolata}

\usepackage{amsmath}
\usepackage{graphicx}
\usepackage{adjustbox}
\usepackage{booktabs}
\usepackage{multicol}
\usepackage{multirow}
\usepackage{makecell}
\usepackage{colortbl}
\usepackage{tabularx}

\usepackage{comment}
\usepackage{placeins}

\usepackage[table]{xcolor}  
\usepackage{adjustbox}      
\usepackage{amssymb}        

\newcommand{\std}[1]{\ensuremath{\,\scriptscriptstyle\pm #1}}
\newcommand{\dn}[3]{\makebox[1.45em][r]{#1}.\makebox[0.42em][l]{#2}\makebox[1.95em][l]{#3}}
\usepackage{xcolor}
\usepackage{xfp}   
\usepackage[most]{tcolorbox}
\usepackage{capt-of}

\definecolor{YesFg}{HTML}{267BB0}
\definecolor{YesBg}{HTML}{EAF2F8}
\definecolor{NoFg}{HTML}{9163AD}
\definecolor{NoBg}{HTML}{F4EFF7}
\definecolor{SlotFg}{HTML}{5A6472}
\newcommand{\slot}[1]{\textcolor{SlotFg}{$\langle$#1$\rangle$}}
\tcbset{promptbox/.style={
    enhanced, breakable=false, colback=white, colframe=black!28, boxrule=0.5pt,
    arc=2pt, left=7pt, right=7pt, top=7pt, bottom=6pt,
    fonttitle=\scriptsize\bfseries\sffamily, coltitle=white, colbacktitle=black!55,
    attach boxed title to top left={xshift=9pt, yshift=-3pt},
    boxed title style={arc=1pt, boxrule=0pt, left=5pt, right=5pt, top=1pt, bottom=1pt}}}
\colorlet{heatpos}{blue}
\colorlet{heatneg}{red}

\newcommand{\heat}[2]{#2}

\title{What Does LALM Post-training Learn? \\
Rethinking Audio Understanding and Reasoning Capabilities}

\author{
  Hua-Hsuan Chen$^{1*}$, Xuanjun Chen$^{1*}$, Wei-Chung Lu$^1$, Chih-Kai Yang$^1$, Yu-Xiang Lin$^1$ \\ \textbf{Sung-Feng Huang}$^2$, \textbf{Haibin Wu}$^3$, \textbf{Hung-yi Lee}$^{1,4}$ \\
 National Taiwan University$^1$ \quad NVIDIA$^2$ \quad Independent Researcher$^3$ \quad NTU AI-Core$^4$ \\
}

\begin{document}

\maketitle
\begingroup
\endgroup


\begin{abstract}
Post-training often improves the performance of large audio-language models on audio understanding and reasoning benchmarks. 
However, these improvements may reflect adaptation to specific question formats, such as multiple-choice questions, rather than genuine advances in audio understanding and reasoning capabilities.
To investigate what LALMs actually learn from post-training, we conduct controlled experiments that disentangle improvements in audio understanding from adaptation to specific question formats. 
Across different models and post-training methods, we find that much of the reported improvement persists even without informative audio, but largely disappears when the question format changes. 
These findings suggest that post-training often teaches models to exploit task-specific textual cues and answer patterns rather than developing generalizable audio understanding and reasoning capabilities. 
\end{abstract}


\section{Introduction}
\label{sec:intro}
Supervised fine-tuning (SFT), rejection fine-tuning (RFT)~\cite{yuan2023rft}, and reinforcement learning methods including  GRPO~\cite{shao2024deepseekmath,guo2025deepseekr1} and VCRL~\cite{jiang2025vcrl} have become standard post-training recipes and have proven effective for LLMs. Recent work has also applied these methods to large audio-language models (LALMs), yielding substantial gains on benchmarks built for audio understanding and reasoning, such as MMAU~\cite{sakshi2024mmau} and MMAR~\cite{ma2025mmar}.
While post-training successfully improves overall performance, recent studies have made a surprising observation: certain gains can be achieved even without audio inputs~\cite{rouditchenko2025omnir1}.
Therefore, it raises a fundamental question: does post-training actually improve audio understanding and reasoning, or does it merely rely on improved text-only capabilities to answer without listening?

\begin{figure}[t]
\centering
\includegraphics[width=\columnwidth]{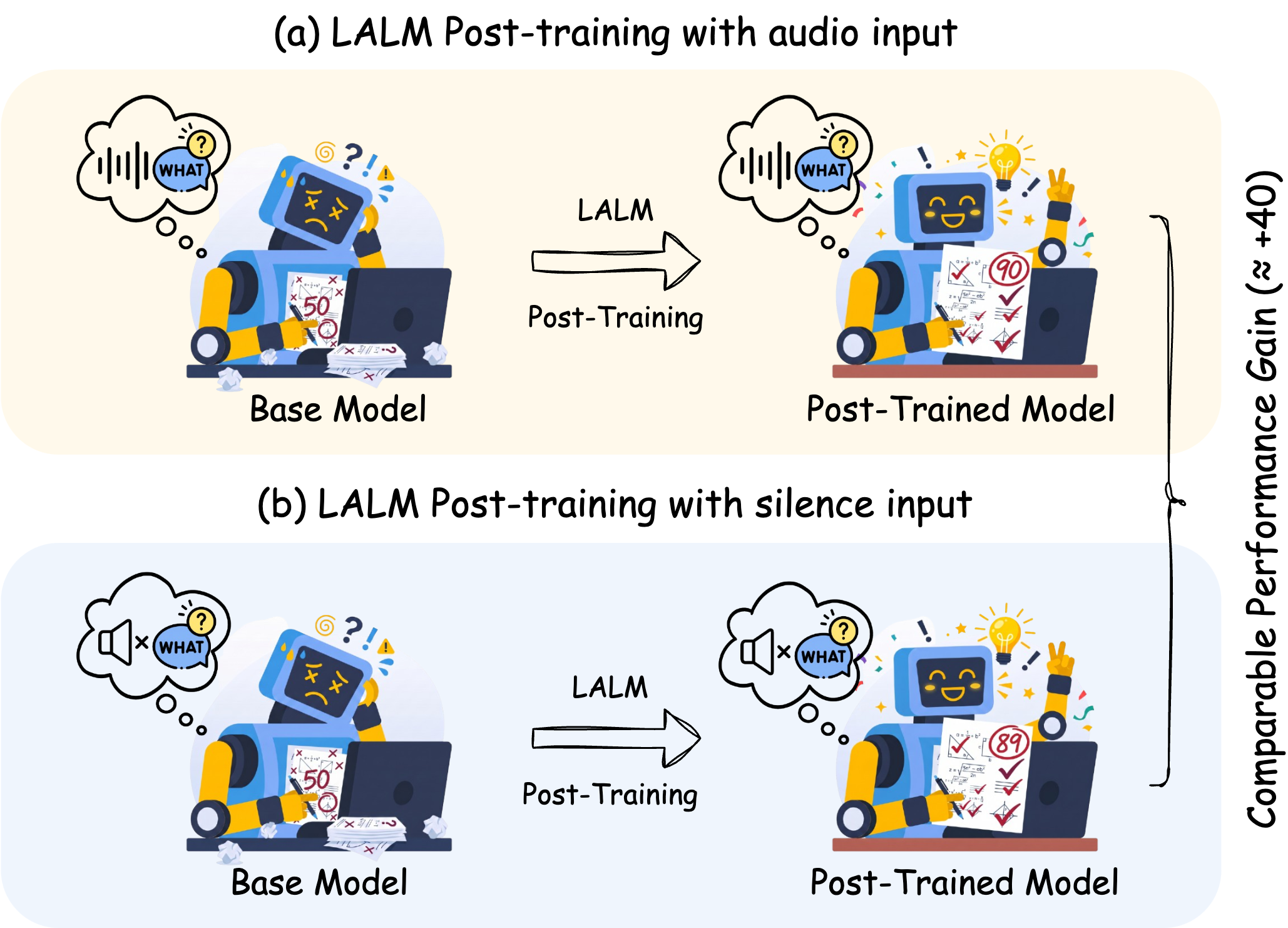}
\caption{Schematic overview of LALM post-training. The illustration demonstrates that models achieve comparable performance gains (e.g., $\sim$+40) whether post-trained with informative audio (a) or silence (b). }
\label{fig:overview}
\end{figure}



To investigate this, we must determine whether models genuinely utilize audio during inference, because a correct answer with audio present is inconclusive due to potential text-only shortcuts. To isolate the true contribution of audio understanding and reasoning, we evaluate both base and post-trained LALM models under different input conditions. As illustrated in Figure~\ref{fig:overview}, our framework compares post-training performance when models are provided with informative audio (a) versus silence input (b). Building upon this controlled evaluation protocol, our study is structured around two core dimensions to thoroughly uncover what models learn. First, we examine the necessity of audio during both evaluation and training by introducing silence controls. Second, since post-training gains may stem from exploiting textual cues rather than acquiring genuine auditory capabilities, we further investigate whether these improvements depend on the multiple-choice format by removing answer options. To answer the overarching question of what LALMs actually learn during post-training, our analysis reveals three main findings:

\begin{itemize}
    \item \textbf{LALM post-training yields comparable performance gains regardless of audio and silence input. }
    Our largest performance gains on MMAU and MMAR ($+14.8$ and $+7.2$ points, respectively) retain $+11.2$ and $+4.9$ points under silence, and models post-trained entirely on silence still improve on real audio.
    \item \textbf{Most LALM performance gains are relied on the text question format.} 
    When evaluated in an open-ended setting, models suffer severe performance degradation, maintaining negligible reported gains despite post-training. 
   \item \textbf{The audio-grounding performance gains of LALM are marginal.} 
   Across different evaluation settings, improvements depending on audio input account for a negligible fraction, calling into question the model's actual audio understanding and reasoning capabilities.
\end{itemize}

\section{Related Work}
\label{sec:related}

\noindent\textbf{Post-training for audio reasoning.}
The post-training recipes now used on audio were developed on text, where their reported benefits were
established on mathematics and code, and they have been carried over largely unchanged.
Chain-of-thought prompting was examined first~\cite{ma2025audiocot}, then reasoning
traces trained at scale~\cite{xie2025audioreasoner}, and reinforcement learning with
verifiable rewards now dominates: answer-level group-relative
optimisation~\cite{li2025r1aqa,rouditchenko2025omnir1,zhao2025keomnir} and variants that
change the curriculum, the reward granularity or the training
data~\cite{wen2025sari,diao2025soundmind,wu2025audiothinker,he2025audiocontribution}.
In all these approaches, the reward function verifies only the final answer, providing no signal on how that answer was derived. Consequently, these studies primarily report overall accuracy, as it is the sole metric optimized during training. In this work, we hold the training setting fixed and investigate whether this accuracy stems from true audio reasoning or mere text shortcuts. 

\noindent\textbf{Whether LALMs listen.}
Unlike text LLMs, LALMs must extract reasoning evidence from an audio waveform rather than text tokens. Since a correct answer can also be derived from the language model's internal knowledge, higher audio-QA accuracy does not guarantee actual audio comprehension. Previous work has found that many benchmark items can be answered with no audio or silent audio~\cite{foo2026glittersaudiorethinkingtext,he2025audiocontribution}, revealed that models follow the text over the acoustic signal when the two disagree~\cite{wang2025audiotextdisagreerevealing}, and shown that strong reasoning scores coexist with weak acoustic perception~\cite{zhang-etal-2026-beyond-transcription}, which decays further as reasoning lengthens~\cite{mao2026scalingfailsmitigatingaudio}. Post-training gains have likewise been traced to text-based reasoning~\cite{rouditchenko2025omnir1}. 
However, whether post-trained LALM models rely on the audio or text capabilities to answer remains unclear. 
Standard evaluations based on absolute accuracy fail to capture this. 
To address this gap, we introduce a controlled evaluation protocol to isolate the true audio contribution from absolute accuracy. 
By doing so, we shed light on what post-training actually learns, offering a principled way to rethink and evaluate audio understanding and reasoning capabilities in LALMs. 


\begin{table*}[t]
\centering
\setlength{\tabcolsep}{6pt}
\begin{adjustbox}{max width=\textwidth}
\begin{tabular}{@{\hspace{4pt}}lcccccc@{\hspace{14pt}}cccccc@{\hspace{4pt}}}
\toprule
& \multicolumn{6}{c}{Qwen2.5-Omni-3B} & \multicolumn{6}{c}{Qwen2-Audio-7B-Instruct} \\
\cmidrule(lr){2-7} \cmidrule(lr){8-13}
& \multicolumn{3}{c}{MMAU} & \multicolumn{3}{c}{MMAR} & \multicolumn{3}{c}{MMAU} & \multicolumn{3}{c}{MMAR} \\
\cmidrule(lr){2-4} \cmidrule(lr){5-7} \cmidrule(lr){8-10} \cmidrule(lr){11-13}
Method & $\mathrm{Acc}$ & $\Delta\mathrm{Acc}^{a}$ & $\Delta\mathrm{Acc}^{s}$ & $\mathrm{Acc}$ & $\Delta\mathrm{Acc}^{a}$ & $\Delta\mathrm{Acc}^{s}$ & $\mathrm{Acc}$ & $\Delta\mathrm{Acc}^{a}$ & $\Delta\mathrm{Acc}^{s}$ & $\mathrm{Acc}$ & $\Delta\mathrm{Acc}^{a}$ & $\Delta\mathrm{Acc}^{s}$ \\
\midrule
\rowcolor{blue!5} Base & 63.1 &    &    & 51.8 &    &    & 45.2 &    &    & 34.7 &    &    \\
SFT & 69.8\std{0.7} & +6.7 & +6.6\std{0.6} & 55.3\std{0.4} & +3.5 & +2.2\std{0.6} & 55.9\std{0.9} & +10.7 & +8.4\std{0.6} & 41.3\std{1.6} & +6.6 & +6.2\std{0.7} \\
RFT & 65.4\std{0.9} & +2.3 & +2.1\std{0.7} & 52.7\std{0.1} & +0.9 & -0.4\std{0.3} & 48.2\std{0.5} & +3.0 & +2.3\std{1.9} & 35.3\std{0.6} & +0.6 & +0.1\std{0.3} \\
GRPO & 71.1\std{0.4} & +8.0 & +6.7\std{1.7} & 55.2\std{0.8} & +3.4 & +0.7\std{0.7} & 60.0\std{1.3} & +14.8 & +11.2\std{0.4} & 41.9\std{0.6} & +7.2 & +4.9\std{0.4} \\
VCRL & 71.1\std{0.6} & +8.0 & +6.6\std{0.3} & 54.6\std{0.9} & +2.8 & +0.5\std{1.7} & 57.6\std{2.1} & +12.4 & +11.0\std{1.4} & 39.2\std{0.7} & +4.5 & +3.6\std{0.8} \\
\bottomrule
\end{tabular}
\end{adjustbox}
\vspace{-4pt}
\caption{Comparison of Audio and Silence Gains across different post-training methods. }
\label{tab:main}
\end{table*}

\begin{table*}[t]
\centering
\small
\setlength{\tabcolsep}{2pt}
\begin{adjustbox}{max width=\textwidth}
\begin{tabular*}{\textwidth}{@{\hspace{4pt}\extracolsep{\fill}}ll ccc ccc@{\hspace{4pt}}}
\toprule
& & \multicolumn{3}{c}{MMAU} & \multicolumn{3}{c}{MMAR} \\
\cmidrule(lr){3-5} \cmidrule(lr){6-8}
Model Backbone & Open-Source Models & $\mathrm{Acc}$ & $\Delta\mathrm{Acc}^{a}$ & $\Delta\mathrm{Acc}^{s}$ & $\mathrm{Acc}$ & $\Delta\mathrm{Acc}^{a}$ & $\Delta\mathrm{Acc}^{s}$ \\
\midrule
Qwen2.5-Omni-3B
& Ke-Omni-R-3B~\cite{zhao2025keomnir}
& 63.1 & +8.6 & +6.6 & 51.8 & +5.4 & +3.0 \\
\midrule
\multirow{3}{*}{Qwen2.5-Omni-7B}
& Ke-Omni-R-7B~\cite{zhao2025keomnir}
& \multirow{3}{*}{71.4} & +4.5 & +1.6
& \multirow{3}{*}{58.7} & +3.9 & +2.8 \\
& AudioMCQ (M$\to$S)~\cite{he2025audiocontribution}
& & +5.0 & +6.4 & & +8.3 & +7.0 \\
& Omni-R1 (AVQA)~\cite{rouditchenko2025omnir1}
& & +4.2 & +1.5 & & +3.1 & +1.0 \\
\midrule
\multirow{2}{*}{Qwen2-Audio-7B-Instruct}
& R1-AQA~\cite{li2025r1aqa}
& \multirow{2}{*}{45.2} & +23.4 & +17.2
& \multirow{2}{*}{34.7} & +17.6 & +10.1 \\
& Audio-Reasoner~\cite{xie2025audioreasoner}
& & +20.5 & +18.4 & & +11.8 & +9.9 \\
\bottomrule
\end{tabular*}
\end{adjustbox}
\vspace{-4pt}
\caption{Comparison of Audio and Silence Gains across open-source models. }
\label{tab:external}
\end{table*}

\section{Controlled Evaluation Protocol}
\label{sec:protocol}


\paragraph{Audio and silence evaluation.}
To isolate the audio-specific improvements of post-training, we evaluate each post-trained model under real audio and same-length silence input conditions, keeping text inputs constant and measuring performance relative to the base model. 
Letting $\mathrm{Acc}^{a}$ and $\mathrm{Acc}^{s}$ denote the accuracies under real audio and silence, we define the performance gains as:
\begin{equation}
\begin{aligned}
\Delta\mathrm{Acc}^{a}
&= \mathrm{Acc}^{a}_{\mathrm{post}}
 - \mathrm{Acc}^{a}_{\mathrm{base}},
 && a:\ \text{audio} \\[2pt]
\Delta\mathrm{Acc}^{s}
&= \mathrm{Acc}^{s}_{\mathrm{post}}
 - \mathrm{Acc}^{s}_{\mathrm{base}},
 && s:\ \text{silence}
\end{aligned}
\label{eq:gains}
\end{equation}
We refer to $\Delta\mathrm{Acc}^{a}$ as Audio Gain, which reflects the overall performance improvement under standard evaluation, and $\Delta\mathrm{Acc}^{s}$ as Silence Gain, which reveals how much of that improvement persists even in the absence of any audio input. 

\paragraph{Multiple-choice and open-ended evaluation.}
We evaluate performance on each benchmark under two answer formats: the original \emph{multiple-choice} setting, where the model's response is checked against provided options, and an \emph{open-ended} evaluation, which removes these options and presents the question independently. 
The model generates a free-form response, which is evaluated by an LLM judge for semantic equivalence to the reference. Since omitting options makes some questions unanswerable, the necessary rewriting rules and judge prompts are provided in Appendix~\ref{app:openend}, with all other evaluation settings remaining unchanged.


\section{Experimental Setting}
\label{sec:setting}


\textbf{Training data. }
The training data pool contains 12{,}267 multiple-choice audio QA items derived from Audio-Cogito~\cite{li2026audiocogitodeepaudioreasoning} (using only the question and gold option). 
These items cover 7{,}129 clips (17.4 hours) across three public corpora, representing distinct domains: 
AudioSet~\cite{jort_audioset_2017} (sound; 3{,}675 clips, 6{,}196 items), 
Medley-solos-DB~\cite{lostanlen2016medley} (music; 2{,}874 clips, 4{,}913 items), 
and DailyTalk~\cite{lee2022dailytalk} (speech; 580 clips, 1{,}158 items).
The pool is split into 10{,}446 training items and a held-out in-domain evaluation set of 1{,}821 items. 
We further verified via audio-level matching that no training audio leaks into the evaluation set.


\textbf{Post-training methods. }
We compare four post-training methods (SFT, RFT~\cite{yuan2023rft}, GRPO~\cite{shao2024deepseekmath}, and VCRL~\cite{jiang2025vcrl}). 
While their specific configurations follow the original papers, they are compared under a unified training budget, sharing a single reward that verifies whether the final answer matches the gold option. 
They also share one recipe (200 LoRA steps on the training pool, three seeds), applied unchanged to two base models: Qwen2.5-Omni-3B~\cite{qwen25omni} and Qwen2-Audio-7B-Instruct~\cite{chu2024qwen2audio}. 
Implementations and hyperparameters are in Appendices~\ref{app:training} and \ref{app:methods}.

\textbf{Evaluation benchmarks and setup. }
We evaluate all models on MMAU test-mini~\cite{sakshi2024mmau} (1{,}000 items) and MMAR~\cite{ma2025mmar} (1{,}000 items). 
During inference, each model uses its own specific prompt and follows a fixed per-benchmark setting shared with its base model. 
Per-capability results rely on the benchmarks' own taxonomy: 
MMAU's \emph{category} and MMAR's \emph{layers}. 
For open-ended evaluation, we use Qwen3.6-27B~\cite{qwen3.6-27b} as an LLM judge to determine whether the generated answers are equivalent to the reference answers. 
Table~\ref{tab:mcvsopen_full} reports all accuracies, which are averaged over three random seeds for our post-trained models at fixed checkpoints.

\section{The Necessity of Audio for LALMs}
\label{sec:results}

Before diving into the specific mechanisms behind the post-training improvements, we first examine a core premise: the necessity of audio for LALMs. In this section, we benchmark model performance against silence controls to assess genuine audio reliance, investigating whether audio is truly required during both evaluation and training phases. 

\begin{table*}[t]
\centering
\small
\setlength{\tabcolsep}{4pt}
\begin{adjustbox}{max width=\textwidth}
\begin{tabularx}{\textwidth}{@{\hspace{4pt}} l >{\hsize=0.96\hsize\centering\arraybackslash}X >{\hsize=1.2\hsize\centering\arraybackslash}X *{4}{>{\hsize=0.96\hsize\centering\arraybackslash}X} @{\hspace{4pt}}}
\toprule
& \multicolumn{2}{c}{MMAU} & \multicolumn{4}{c}{MMAR} \\
\cmidrule(lr){2-3} \cmidrule(lr){4-7}
& Reasoning & Info Extraction & Signal & Perception & Semantic & Cultural \\
\midrule
\rowcolor{gray!15} \multicolumn{7}{@{}l}{\textit{Qwen2.5-Omni-3B}} \\[2pt]
\rowcolor{blue!5} Base & \dn{43}{1}{} & \dn{46}{1}{} & \dn{44}{2}{} & \dn{31}{2}{} & \dn{36}{2}{} & \dn{29}{1}{} \\
SFT & \heat{5.7}{\dn{+5}{7}{\std{0.9}}} & \heat{8.6}{\dn{+8}{6}{\std{0.4}}} & \heat{-0.8}{\dn{-0}{8}{\std{2.7}}} & \heat{0.7}{\dn{+0}{7}{\std{0.6}}} & \heat{4.0}{\dn{+4}{0}{\std{0.7}}} & \heat{2.6}{\dn{+2}{6}{\std{3.0}}} \\
RFT & \heat{3.0}{\dn{+3}{0}{\std{0.9}}} & \heat{0.3}{\dn{+0}{3}{\std{1.3}}} & \heat{-3.1}{\dn{-3}{1}{\std{4.8}}} & \heat{0.1}{\dn{+0}{1}{\std{0.4}}} & \heat{-0.2}{\dn{-0}{2}{\std{0.6}}} & \heat{-1.2}{\dn{-1}{2}{\std{1.1}}} \\
GRPO & \heat{6.2}{\dn{+6}{2}{\std{1.9}}} & \heat{7.8}{\dn{+7}{8}{\std{1.8}}} & \heat{-3.9}{\dn{-3}{9}{\std{4.8}}} & \heat{-0.2}{\dn{-0}{2}{\std{0.1}}} & \heat{1.5}{\dn{+1}{5}{\std{0.8}}} & \heat{2.1}{\dn{+2}{1}{\std{4.4}}} \\
VCRL & \heat{6.2}{\dn{+6}{2}{\std{0.1}}} & \heat{7.5}{\dn{+7}{5}{\std{1.2}}} & \heat{1.6}{\dn{+1}{6}{\std{11.0}}} & \heat{-0.8}{\dn{-0}{8}{\std{0.1}}} & \heat{1.2}{\dn{+1}{2}{\std{2.3}}} & \heat{1.9}{\dn{+1}{9}{\std{1.8}}} \\
Ke-Omni-R-3B$^*$ & \heat{7.8}{\dn{+7}{8}{}} & \heat{4.0}{\dn{+4}{0}{}} & \heat{2.3}{\dn{+2}{3}{}} & \heat{3.0}{\dn{+3}{0}{}} & \heat{2.4}{\dn{+2}{4}{}} & \heat{5.0}{\dn{+5}{0}{}} \\
\midrule
\rowcolor{gray!15} \multicolumn{7}{@{}l}{\textit{Qwen2.5-Omni-7B}} \\[2pt]
\rowcolor{blue!5} Base & \dn{52}{1}{} & \dn{56}{7}{} & \dn{55}{8}{} & \dn{36}{4}{} & \dn{40}{8}{} & \dn{35}{5}{} \\
Ke-Omni-R-7B$^*$ & \heat{1.0}{\dn{+1}{0}{}} & \heat{2.8}{\dn{+2}{8}{}} & \heat{-4.7}{\dn{-4}{7}{}} & \heat{3.5}{\dn{+3}{5}{}} & \heat{3.9}{\dn{+3}{9}{}} & \heat{0.0}{\dn{0}{0}{}} \\
AudioMCQ (M$\to$S)$^*$ & \heat{6.2}{\dn{+6}{2}{}} & \heat{6.8}{\dn{+6}{8}{}} & \heat{-7.0}{\dn{-7}{0}{}} & \heat{9.4}{\dn{+9}{4}{}} & \heat{9.5}{\dn{+9}{5}{}} & \heat{-2.8}{\dn{-2}{8}{}} \\
Omni-R1 (AVQA)$^*$ & \heat{2.7}{\dn{+2}{7}{}} & \heat{-0.9}{\dn{-0}{9}{}} & \heat{2.3}{\dn{+2}{3}{}} & \heat{0.0}{\dn{0}{0}{}} & \heat{2.7}{\dn{+2}{7}{}} & \heat{-1.4}{\dn{-1}{4}{}} \\
\midrule
\rowcolor{gray!15} \multicolumn{7}{@{}l}{\textit{Qwen2-Audio-7B-Instruct}} \\[2pt]
\rowcolor{blue!5} Base & \dn{28}{4}{} & \dn{30}{0}{} & \dn{30}{2}{} & \dn{22}{3}{} & \dn{27}{7}{} & \dn{26}{2}{} \\
SFT & \heat{9.0}{\dn{+9}{0}{\std{0.6}}} & \heat{7.3}{\dn{+7}{3}{\std{0.9}}} & \heat{0.0}{\dn{0}{0}{\std{0.0}}} & \heat{7.8}{\dn{+7}{8}{\std{0.9}}} & \heat{6.3}{\dn{+6}{3}{\std{0.9}}} & \heat{2.8}{\dn{+2}{8}{\std{1.2}}} \\
RFT & \heat{2.4}{\dn{+2}{4}{\std{1.6}}} & \heat{2.2}{\dn{+2}{2}{\std{2.6}}} & \heat{-3.1}{\dn{-3}{1}{\std{3.6}}} & \heat{0.3}{\dn{+0}{3}{\std{0.8}}} & \heat{0.9}{\dn{+0}{9}{\std{0.1}}} & \heat{-1.7}{\dn{-1}{7}{\std{0.4}}} \\
GRPO & \heat{12.9}{\dn{+12}{9}{\std{0.8}}} & \heat{7.5}{\dn{+7}{5}{\std{0.5}}} & \heat{-3.9}{\dn{-3}{9}{\std{4.8}}} & \heat{6.8}{\dn{+6}{8}{\std{0.4}}} & \heat{6.1}{\dn{+6}{1}{\std{0.4}}} & \heat{-1.9}{\dn{-1}{9}{\std{1.5}}} \\
VCRL & \heat{12.1}{\dn{+12}{1}{\std{1.6}}} & \heat{8.7}{\dn{+8}{7}{\std{1.1}}} & \heat{-3.9}{\dn{-3}{9}{\std{3.6}}} & \heat{4.8}{\dn{+4}{8}{\std{0.3}}} & \heat{5.3}{\dn{+5}{3}{\std{1.1}}} & \heat{-2.8}{\dn{-2}{8}{\std{0.7}}} \\
R1-AQA$^*$ & \heat{17.7}{\dn{+17}{7}{\std{0.4}}} & \heat{16.0}{\dn{+16}{0}{\std{0.2}}} & \heat{14.0}{\dn{+14}{0}{\std{0.0}}} & \heat{12.6}{\dn{+12}{6}{\std{0.2}}} & \heat{9.6}{\dn{+9}{6}{\std{0.1}}} & \heat{2.8}{\dn{+2}{8}{\std{0.0}}} \\
Audio-Reasoner$^*$ & \heat{14.0}{\dn{+14}{0}{}} & \heat{27.6}{\dn{+27}{6}{}} & \heat{9.3}{\dn{+9}{3}{}} & \heat{7.9}{\dn{+7}{9}{}} & \heat{12.9}{\dn{+12}{9}{}} & \heat{7.1}{\dn{+7}{1}{}} \\
\bottomrule
\end{tabularx}
\end{adjustbox}
\vspace{-4pt}
\caption{Comparison of Silence Gains across different audio capabilities. $^*$Denotes open-source models. }
\label{tab:cap_listen}
\end{table*}

\subsection{Evaluating Overall Gains Across Models}
\label{ssec:effective}

\textbf{Our post-trained models. } Table~\ref{tab:main} compares the Audio and Silence Gains of the four post-training methods across our base models. 
Ideally, post-training should yield a substantial Audio Gain ($\Delta\mathrm{Acc}^{a}$) with near-zero Silence Gain ($\Delta\mathrm{Acc}^{s}$), reflecting genuine reliance on audio. 
However, the results reveal that these two metrics are highly correlated. 
For instance, while GRPO on Qwen2-Audio-7B-Instruct improves MMAU accuracy by $14.8$ points with real audio, a striking $11.2$-point gain is observed even under silence. 
In contrast, when the Silence Gain is near zero (e.g., RFT on Qwen2.5-Omni-3B), Audio Gain is a marginal $0.9$\%. 

\textbf{Open-source models. } One might argue that this observed phenomenon stems from our specific training recipe rather than post-training in general. To address this, Table~\ref{tab:external} applies the same silence control to several existing open-source models, each evaluated using the specific prompt and sampling strategy detailed in its respective paper or repository. 
The results reveal an identical pattern. Although their overall accuracy gains are often larger than those of our models, the majority of these improvements persist under silence. For example, R1-AQA and Audio-Reasoner (both built on the same base model) achieve Audio Gains of $23.4$ and $20.5$ points on MMAU, respectively, yet they still exhibit substantial Silence Gains of $17.2$ and $18.4$ points. Even more strikingly, AudioMCQ, which specifically trains on a subset of items filtered for strong audio reliance~\cite{he2025audiocontribution}, yields an Audio Gain of $5.0$ points on MMAU but a Silence Gain of $6.4$ points, meaning the improvement is actually higher without the real audio. 


Across both Tables~\ref{tab:main} and~\ref{tab:external}, Audio Gains and Silence Gains remain closely aligned for almost every model. Ultimately, this consistent coupling indicates that these post-training recipes do not teach the models to listen. Instead, they largely amplify the models' reliance on textual artifacts and guessing strategies, rendering the apparent audio improvements somewhat illusory.

\subsection{Evaluating Fine-Grained Capabilities}
\label{ssec:listening}


To determine whether this reliance on textual shortcuts is universal or task-specific, we conduct a fine-grained analysis across different audio capabilities. 
Table~\ref{tab:cap_listen} details the Silence Gains ($\Delta\mathrm{Acc}^{s}$) for specific capability categories. 
On MMAU, both reasoning and information extraction show substantial Silence Gains, typically around seven points for the stronger methods, with some open-source models exhibiting even larger gains. 
MMAR displays greater variance: semantic and perception capabilities retain positive Silence Gains for most models, whereas cultural gains are generally smaller, and the signal layer fluctuates widely from $-7.0$ to $+14.0$ points. 
Despite these variations, the overarching pattern holds: across multiple fine-grained audio capabilities, Silence Gains frequently deviate from the expected near-zero baseline. 
Ultimately, these findings confirm that much of the apparent performance improvement persists even when audio is removed during evaluation. 

\begin{table}[t]
\centering
\small
\setlength{\tabcolsep}{0pt}
\begin{adjustbox}{max width=\columnwidth}
\begin{tabularx}{\columnwidth}{@{\hspace{4pt}} l *{4}{>{\centering\arraybackslash}X} @{\hspace{4pt}}}
\toprule
& \multicolumn{2}{c}{MMAU} & \multicolumn{2}{c}{MMAR} \\
\cmidrule(lr){2-3} \cmidrule(lr){4-5}
Method & Audio & Silence & Audio & Silence \\
\midrule
\rowcolor{gray!15} \multicolumn{5}{@{\hspace{4pt}}l}{\textit{Qwen2.5-Omni-3B}} \\[2pt]
\rowcolor{blue!5} Base & 63.1 &  & 51.8 &  \\
SFT   & +6.7 & +3.8 & +3.5 & +0.5 \\
RFT   & +2.3 & +1.0 & +0.9 & -0.2 \\
GRPO  & +8.0 & +4.0 & +3.4 & +1.0 \\
VCRL  & +8.0 & +3.1 & +2.8 & -0.6 \\
\midrule
\rowcolor{gray!15} \multicolumn{5}{@{\hspace{4pt}}l}{\textit{Qwen2-Audio-7B-Instruct}} \\[2pt]
\rowcolor{blue!5} Base & 45.2 &    & 34.7 &    \\
SFT   & +10.7 & +7.5 & +6.6 & +4.4 \\
RFT   & +3.0 & +5.3 & +0.6 & +1.6 \\
GRPO  & +14.8 & +7.6 & +7.2 & +2.1 \\
VCRL  & +12.4 & +6.3 & +4.5 & +0.2 \\
\bottomrule
\end{tabularx}
\end{adjustbox}
\vspace{-4pt}
\caption{Audio Gains of models trained on real audio versus silence, evaluated under the real-audio condition.}
\label{tab:blind}
\end{table}

\subsection{Post-Training with Audio versus Silence}
\label{ssec:blind}

While evaluating with silence demonstrates that performance gains persist without audio, it remains unclear whether audio was strictly necessary to acquire these capabilities during training. To isolate the role of training audio, we replicate the post-training process by replacing all audio inputs with same-length silence, examining how much of the performance improvement is retained when the model is subsequently evaluated on real audio. 

Table~\ref{tab:blind} shows that on MMAU the silence-trained models keep about half of the gain: GRPO on Qwen2-Audio-7B-Instruct adds $14.8$ points trained on real audio and still $7.6$ trained on nothing but silence. A model that heard no information at all for the whole of post-training arrives at half the improvement, on the same benchmark, decoded from the same real audio. 
MMAR retains less of its gain and exhibits greater variance, indicating a stronger reliance on training audio. 




\section{What Does LALM Post-training Learn?}
\label{sec:open}

The preceding analysis demonstrates that much of the post-training improvement persists even without informative audio. This naturally leads to the central question of our work: if not genuine audio understanding, what do these models actually learn? We hypothesize that rather than acquiring new auditory capabilities, the models are instead learning to exploit textual cues inherent in the multiple-choice format. In this section, we answer this question by systematically isolating the impact of answer options, analyzing how model behavior shifts when the multiple-choice format is removed from both training and evaluation.

\begin{table}[t]
\centering
\small
\setlength{\tabcolsep}{0pt}
\begin{tabularx}{\columnwidth}{@{\hspace{3pt}} l *{6}{>{\centering\arraybackslash}X} @{\hspace{3pt}}}
\toprule
& \multicolumn{3}{c}{MMAU} & \multicolumn{3}{c}{MMAR} \\
\cmidrule(lr){2-4} \cmidrule(lr){5-7}
& Gold & Listed & Invalid & Gold & Listed & Invalid \\
\midrule
\rowcolor{gray!15} \multicolumn{7}{@{}l}{\textit{Omni-3B}} \\[2pt]
\rowcolor{blue!5} Base & 63.1 & 29.4 & 7.5 & 51.8 & 45.0 & 3.2 \\
SFT & 69.8 & 28.6 & 1.6 & 55.3 & 43.5 & 1.2 \\
RFT & 65.4 & 29.0 & 5.6 & 52.7 & 44.1 & 3.2 \\
GRPO & 71.1 & 26.9 & 2.0 & 55.2 & 42.9 & 1.8 \\
VCRL & 71.1 & 27.0 & 1.9 & 54.6 & 44.0 & 1.4 \\
Ke-Omni-R$^*$ & 71.7 & 27.6 & 0.7 & 57.2 & 42.4 & 0.4 \\
\midrule
\rowcolor{gray!15} \multicolumn{7}{@{}l}{\textit{Omni-7B}} \\[2pt]
\rowcolor{blue!5} Base & 71.4 & 28.2 & 0.4 & 58.7 & 40.6 & 0.7 \\
Ke-Omni-R$^*$ & 75.9 & 23.5 & 0.6 & 62.6 & 37.3 & 0.1 \\
AudioMCQ$^*$ & 76.4 & 23.1 & 0.5 & 67.0 & 32.6 & 0.4 \\
Omni-R1$^*$ & 75.6 & 24.1 & 0.3 & 61.8 & 37.7 & 0.5 \\
\midrule
\rowcolor{gray!15} \multicolumn{7}{@{}l}{\textit{Qwen2-Audio}} \\[2pt]
\rowcolor{blue!5} Base & 45.2 & 37.6 & 17.2 & 34.7 & 50.5 & 14.8 \\
SFT & 55.9 & 35.6 & 8.5 & 41.3 & 51.2 & 7.5 \\
RFT & 48.2 & 35.8 & 16.1 & 35.3 & 50.8 & 13.9 \\
GRPO & 60.0 & 32.4 & 7.6 & 41.9 & 50.6 & 7.5 \\
VCRL & 57.6 & 32.9 & 9.5 & 39.2 & 51.4 & 9.4 \\
R1-AQA$^*$ & 68.6 & 30.4 & 1.0 & 52.3 & 46.7 & 1.0 \\
Audio-Reas.$^*$ & 65.7 & 30.3 & 4.0 & 46.5 & 48.1 & 5.4 \\
\bottomrule
\end{tabularx} 
\vspace{-4pt}
\caption{Answer Distribution across Multiple-choice Options. $^*$Denotes open-source model.}
\label{tab:answerset}
\end{table}

\begin{table*}[t]
\centering
\setlength{\tabcolsep}{4pt}
\begin{adjustbox}{max width=\textwidth}
\begin{tabular}{@{\hspace{4pt}}lcccccc cccccc@{\hspace{4pt}}}
\toprule
& \multicolumn{6}{c}{MMAU} & \multicolumn{6}{c}{MMAR} \\
\cmidrule(lr){2-7} \cmidrule(lr){8-13}
& \multicolumn{3}{c}{\textsc{mc}} & \multicolumn{3}{c}{\textsc{open}} & \multicolumn{3}{c}{\textsc{mc}} & \multicolumn{3}{c}{\textsc{open}} \\
\cmidrule(lr){2-4} \cmidrule(lr){5-7} \cmidrule(lr){8-10} \cmidrule(lr){11-13}
Method & $\mathrm{Acc}$ & $\Delta\mathrm{Acc}^{a}$ & $\Delta\mathrm{Acc}^{s}$ & $\mathrm{Acc}$ & $\Delta\mathrm{Acc}^{a}$ & $\Delta\mathrm{Acc}^{s}$ & $\mathrm{Acc}$ & $\Delta\mathrm{Acc}^{a}$ & $\Delta\mathrm{Acc}^{s}$ & $\mathrm{Acc}$ & $\Delta\mathrm{Acc}^{a}$ & $\Delta\mathrm{Acc}^{s}$ \\
\midrule
\rowcolor{gray!15} \multicolumn{13}{@{}l}{\textit{Qwen2.5-Omni-3B}} \\[2pt]
\rowcolor{blue!5} Base & 63.1 &    &    & 27.1 &    &    & 51.8 &    &    & 28.5 &    &    \\
SFT & 69.8 & +6.7 & +6.6 & 28.8 & +1.7 & -0.7 & 55.3 & +3.5 & +2.2 & 29.0 & +0.5 & +0.1 \\
RFT & 65.4 & +2.3 & +2.1 & 27.7 & +0.7 & -0.3 & 52.7 & +0.9 & -0.4 & 28.6 & +0.1 & -0.7 \\
GRPO & 71.1 & +8.0 & +6.7 & 28.2 & +1.2 & +0.5 & 55.2 & +3.4 & +0.7 & 29.1 & +0.6 & -0.1 \\
VCRL & 71.1 & +8.0 & +6.6 & 28.4 & +1.4 & -0.5 & 54.6 & +2.8 & +0.5 & 28.8 & +0.3 & -0.4 \\
Ke-Omni-R-3B$^*$ & 71.7 & +8.6 & +6.6 & 28.2 & +1.1 & +0.6 & 57.2 & +5.4 & +3.0 & 28.4 & -0.1 & +0.2 \\
\midrule
\rowcolor{gray!15} \multicolumn{13}{@{}l}{\textit{Qwen2.5-Omni-7B}} \\[2pt]
\rowcolor{blue!5} Base & 71.4 &    &    & 31.1 &    &    & 58.7 &    &    & 31.2 &    &    \\
Ke-Omni-R-7B$^*$ & 75.9 & +4.5 & +1.6 & 32.9 & +1.8 & +2.5 & 62.6 & +3.9 & +2.8 & 33.6 & +2.4 & -1.2 \\
AudioMCQ (M$\to$S)$^*$ & 76.4 & +5.0 & +6.4 & 31.2 & +0.1 & +1.6 & 67.0 & +8.3 & +7.0 & 32.4 & +1.2 & -0.8 \\
Omni-R1 (AVQA)$^*$ & 75.6 & +4.2 & +1.5 & 30.9 & -0.2 & +1.1 & 61.8 & +3.1 & +1.0 & 33.1 & +1.9 & +0.4 \\
\midrule
\rowcolor{gray!15} \multicolumn{13}{@{}l}{\textit{Qwen2-Audio-7B-Instruct}} \\[2pt]
\rowcolor{blue!5} Base & 45.2 &    &    & 27.1 &    &    & 34.7 &    &    & 25.1 &    &    \\
SFT & 55.9 & +10.7 & +8.4 & 28.2 & +1.2 & +0.7 & 41.3 & +6.6 & +6.2 & 26.0 & +0.9 & +0.9 \\
RFT & 48.2 & +3.0 & +2.3 & 27.9 & +0.9 & -0.3 & 35.3 & +0.6 & +0.1 & 25.5 & +0.4 & -0.1 \\
GRPO & 60.0 & +14.8 & +11.2 & 29.3 & +2.3 & +0.8 & 41.9 & +7.2 & +4.9 & 26.8 & +1.7 & +1.5 \\
VCRL & 57.6 & +12.4 & +11.0 & 29.2 & +2.1 & +0.4 & 39.2 & +4.5 & +3.6 & 27.0 & +1.9 & +1.6 \\
R1-AQA$^*$ & 68.6 & +23.4 & +17.2 & 26.2 & -0.9 & +0.4 & 52.3 & +17.6 & +10.1 & 25.5 & +0.4 & +1.4 \\
Audio-Reasoner$^*$ & 65.7 & +20.5 & +18.4 & 28.4 & +1.3 & +2.0 & 46.5 & +11.8 & +9.9 & 26.8 & +1.7 & -1.5 \\
\bottomrule
\end{tabular}
\end{adjustbox}
\vspace{-4pt}
\caption{Comparison of Audio and Silence Gains evaluated under the multiple-choice (\textsc{mc}) and open-ended (\textsc{open}) evaluation. Reported accuracies correspond to the post-trained models. $^*$Denotes open-source models.}
\label{tab:mcvsopen_full}
\end{table*}

\subsection{Analyzing shifts in answer patterns}
\label{ssec:answer_d}

To investigate whether models are simply learning to exploit the task format, we first examine how post-training alters their answering patterns under the original multiple-choice setting. A wrong answer can either select an incorrect option or fall outside the provided options entirely (an invalid response). 
Table~\ref{tab:answerset} reports the proportions
of correct answers, incorrect listed options, and
invalid responses before and after post-training.

The proportion of invalid responses decreases
substantially after post-training, while incorrect
listed answers change less. On MMAU,
Qwen2-Audio-7B-Instruct produces answers outside
the option list on 17.2\% of items before post-training,
which falls to 7.6\% after GRPO.
Meanwhile, incorrect listed answers decrease
from 37.6\% to 32.4\%.
A similar pattern appears for Qwen2.5-Omni-3B,
where invalid responses decrease from 7.5\% to
2.0\% after GRPO. These results suggest that part of the improvement
may come from learning to follow the multiple-choice
format and take advantage of textual cues, rather than better
distinguishing among the answer options based on audio, so we next examine whether these gains persist when the answer option list is removed from question.

\subsection{Assessing open-ended evaluation}
\label{ssec:openeval}

To further examine whether these gains depend on the multiple-choice format, we evaluate models in an open-ended setting where they must respond in their own words (details in Appendix~\ref{app:openend}). As shown in Table~\ref{tab:mcvsopen_full}, absolute accuracy drops substantially across the board (e.g., Qwen2.5-Omni-3B drops from $63.1$ to $27.1$ on multiple-choice MMAU). However, since these models and benchmarks were intrinsically designed for multiple-choice, this absolute decrease is expected. Therefore, rather than absolute performance, our analysis focuses on the post-training gains relative to each model's base model under the same evaluation protocol.

Crucially, removing the answer options sharply diminishes both Audio and Silence Gains across all evaluated models and benchmarks. On MMAU, for instance, Audio Gains for GRPO and VCRL (applied to Qwen2-Audio-7B-Instruct) drop from $+14.8$ and $+12.4$ to merely $+2.3$ and $+2.1$ points, respectively. Silence Gains shrink similarly, with GRPO's $\Delta\mathrm{Acc}^{s}$ falling from $+11.2$ to $+0.8$. The open-source models exhibit an equally drastic decline: Audio-Reasoner's Audio Gain plummets from $+20.5$ to $+1.3$, while R1-AQA drops from $+23.4$ to a negative gain of $-0.9$.

Together, these results show that most post-training gains largely fail to transfer beyond the multiple-choice format. This strongly suggests that the observed improvements stem from the models exploiting cues in the textual questions and answer options, rather than reflecting a genuine, broad advancement in audio understanding and reasoning. 

\begin{table*}[t]
\centering
\setlength{\tabcolsep}{6pt}
\begin{adjustbox}{max width=\textwidth}
\begin{tabular}{@{\hspace{4pt}}lcccccc cccccc@{\hspace{4pt}}}
\toprule
& \multicolumn{6}{c}{Qwen2.5-Omni-3B} & \multicolumn{6}{c}{Qwen2-Audio-7B-Instruct} \\
\cmidrule(lr){2-7} \cmidrule(lr){8-13}
& \multicolumn{3}{c}{MMAU} & \multicolumn{3}{c}{MMAR} & \multicolumn{3}{c}{MMAU} & \multicolumn{3}{c}{MMAR} \\
\cmidrule(lr){2-4} \cmidrule(lr){5-7} \cmidrule(lr){8-10} \cmidrule(lr){11-13}
Method & $\mathrm{Acc}$ & $\Delta\mathrm{Acc}^{a}$ & $\Delta\mathrm{Acc}^{s}$ & $\mathrm{Acc}$ & $\Delta\mathrm{Acc}^{a}$ & $\Delta\mathrm{Acc}^{s}$ & $\mathrm{Acc}$ & $\Delta\mathrm{Acc}^{a}$ & $\Delta\mathrm{Acc}^{s}$ & $\mathrm{Acc}$ & $\Delta\mathrm{Acc}^{a}$ & $\Delta\mathrm{Acc}^{s}$ \\
\midrule
\rowcolor{blue!5} Base & 27.0 &    &    & 28.5 &    &    & 27.1 &    &    & 25.1 &    &    \\
SFT & 27.9\std{0.9} & +0.9\std{0.9} & -0.2\std{0.6} & 30.2\std{0.4} & +1.7\std{0.4} & -0.6\std{0.3} & 26.1\std{1.5} & -1.0\std{1.5} & +1.3\std{0.2} & 24.6\std{0.5} & -0.5\std{0.5} & -0.1\std{0.3} \\
RFT & 27.8\std{0.2} & +0.8\std{0.2} & -0.1\std{0.5} & 28.6\std{0.6} & +0.1\std{0.6} & -0.7\std{0.3} & 28.5\std{0.5} & +1.4\std{0.5} & +0.3\std{0.1} & 25.3\std{0.1} & +0.2\std{0.1} & +0.3\std{0.3} \\
GRPO & 28.9\std{0.7} & +1.9\std{0.7} & +0.1\std{0.6} & 29.2\std{1.1} & +0.7\std{1.1} & -1.2\std{0.2} & 29.1\std{1.4} & +2.0\std{1.4} & +0.6\std{0.1} & 26.4\std{0.7} & +1.3\std{0.7} & +0.5\std{0.2} \\
VCRL & 28.7\std{1.2} & +1.7\std{1.2} & -0.5\std{0.7} & 28.6\std{0.3} & +0.1\std{0.3} & -0.5\std{0.6} & 30.1\std{0.9} & +3.0\std{0.9} & +0.7\std{0.5} & 25.9\std{0.3} & +0.8\std{0.3} & 0.0\std{0.9} \\
\bottomrule
\end{tabular}
\end{adjustbox}
\vspace{-4pt}
\caption{Comparison of Audio and Silence Gains following open-ended post-training and open-ended evaluation. }
\label{tab:opentrain}
\end{table*}

\subsection{Examining open-ended training}
\label{ssec:opentrain}

As shown in Section~\ref{ssec:openeval}, evaluating models without multiple-choice options eliminates most post-training gains. However, this sharp decline might simply stem from a train-eval format mismatch (i.e., training with options but evaluating without them). To test this hypothesis, we eliminate the mismatch by repeating the training procedure under a strictly open-ended setting. We remove the option list from all training questions and apply the rewriting and matching rules detailed in Appendix~\ref{app:opentrain}.

Table~\ref{tab:opentrain} reveals that aligning the training format with the open-ended evaluation does not recover the lost gains. For instance, applying open-ended GRPO to Qwen2-Audio-7B-Instruct yields a marginal MMAU Audio Gain of $+2.0$. This is virtually identical to the $+2.3$ gain obtained under mismatched training, and remains far below the $+14.8$ gain observed in the original multiple-choice setting. VCRL follows the same trajectory, reaching an Audio Gain of only $+3.0$. Silence Gains similarly plateau, with GRPO's $\Delta\mathrm{Acc}^{s}$ hovering at merely $+0.6$ after open-ended training.

These results confirm that the performance collapse is not an artifact of format mismatch. Even when explicitly trained to generate answers in their own words, models fail to recover the substantial gains seen under multiple-choice evaluation, reinforcing that these post-training improvements heavily rely on the presence of answer options.

\subsection{Tracing individual error transitions}
\label{ssec:AnsErr}

To examine individual outcomes, we isolate items where the post-trained model successfully corrects the base model. We then re-evaluate these items under silence (\textsc{sil}) and open-ended (\textsc{open}) controls. As Table~\ref{tab:transition} outlines, an outcome is recorded as $\checkmark$ (correct) if only the post-trained model remains correct, and $\times$ (wrong) if both fail; cases where the base model becomes correct are excluded. The four resulting combinations reveal the underlying drivers of the observed corrections: 
\begin{itemize}
    \item \textbf{Ideal} \textit{Wrong in silence, correct open-ended.} The correction genuinely requires the underlying audio and generalizes beyond options. 
    \item \textbf{Spurious} \textit{Correct in silence, wrong open-ended.} The correction ignores audio but relies on explicitly provided answer options. 
    \item \textbf{Artifact} \textit{Correct under both controls.} The correction requires neither audio nor options.
    \item \textbf{Option-bound Audio} \textit{Wrong under both controls.} The correction is sensitive to audio, but the audio is only useful as a secondary cue for selecting among provided textual options. 
\end{itemize}
Table~\ref{tab:transition} shows that these outcomes are distributed very unevenly. The \textbf{Ideal} case is exceptionally rare, peaking at just $7$ items in any setting. In contrast, \textbf{Spurious} corrections are far more common (up to $48$ items). However, the vast majority of original corrections simply fail under both controls (\textbf{Option-bound Audio}). Ultimately, different post-training methods rarely yield robust audio understanding; its apparent gains rely fundamentally on the explicit text-only answer options.


\begin{table}[t]
\centering
\small
\setlength{\tabcolsep}{0pt}
\begin{adjustbox}{max width=\columnwidth}
\begin{tabular*}{\columnwidth}{@{\hspace{3pt}\extracolsep{\fill}}lcccccccc@{\hspace{3pt}}}
\toprule
& \multicolumn{4}{c}{MMAU} & \multicolumn{4}{c}{MMAR} \\
\cmidrule(lr){2-5} \cmidrule(lr){6-9}
\textsc{silence} & $\checkmark$ & $\times$ & $\checkmark$ & $\times$ & $\checkmark$ & $\times$ & $\checkmark$ & $\times$ \\
\textsc{open-ended} & $\times$ & $\checkmark$ & $\checkmark$ & $\times$ & $\times$ & $\checkmark$ & $\checkmark$ & $\times$ \\
\midrule
\rowcolor{gray!15} \multicolumn{9}{@{}l}{\textit{Omni-3B}} \\[2pt]
SFT & 18 & 3 & 1 & 39 & 16 & 0 & 0 & 28 \\
RFT & 10 & 3 & 0 & 26 & 1 & 0 & 0 & 17 \\
GRPO & 24 & 3 & 1 & 38 & 15 & 1 & 0 & 33 \\
VCRL & 20 & 3 & 2 & 36 & 15 & 1 & 2 & 28 \\
\midrule
\rowcolor{gray!15} \multicolumn{9}{@{}l}{\textit{Qwen2-Audio}} \\[2pt]
SFT & 36 & 7 & 3 & 55 & 32 & 2 & 1 & 41 \\
RFT & 6 & 3 & 0 & 28 & 5 & 1 & 0 & 17 \\
GRPO & 48 & 7 & 5 & 62 & 43 & 3 & 1 & 48 \\
VCRL & 40 & 7 & 5 & 59 & 39 & 2 & 2 & 42 \\
\bottomrule
\end{tabular*}
\end{adjustbox}
\vspace{-4pt}
\caption{Correction Outcomes under Silence and Open-ended Controls ($\checkmark$: corrected, $\times$: still wrong).}
\label{tab:transition}
\end{table}

\subsection{Case Study: Revealing option priors}
\label{ssec:case}

\begin{figure*}[t]
  \centering
  \includegraphics[width=0.91\textwidth]{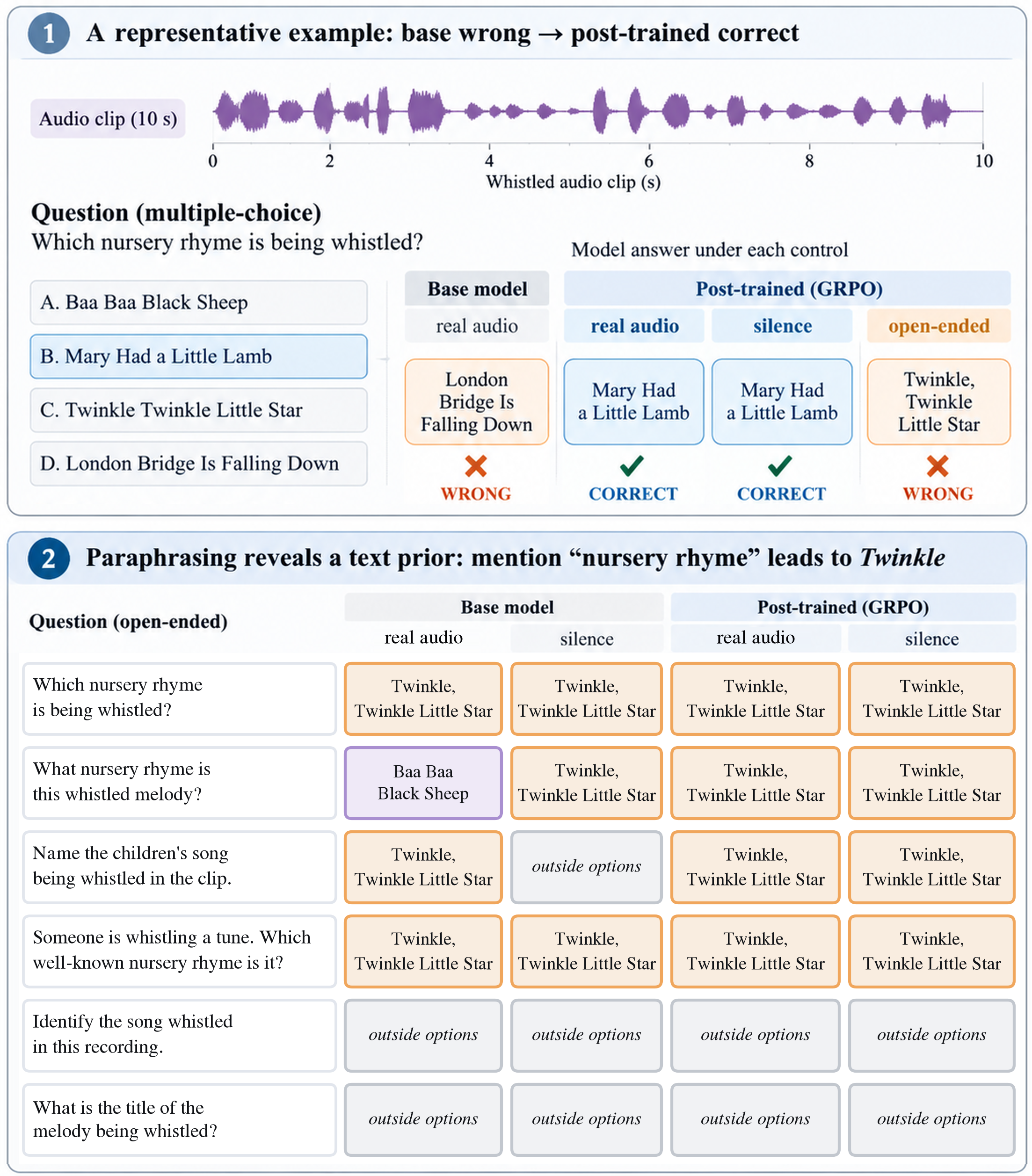}
  \vspace{-4pt}
  \caption{Case Study: Revealing option priors and the impact of question phrasing on model predictions.}
  \label{fig:case}
\end{figure*}

Figure~\ref{fig:case} examines a representative case from MMAU to demonstrate how models exploit textual options to bypass audio reasoning. 
A ten-second audio of whistling is played, accompanied by four nursery rhyme options. The base model answers incorrectly \emph{London Bridge Is Falling Down}, while the GRPO post-trained model answers correctly \emph{Mary Had a Little Lamb}. Under the standard evaluation protocol, this item is recorded as a successful correction and contributes to the Audio Gain reported in Table~\ref{tab:main}.
However, our controls reveal a different story. When the audio is replaced by silence, the GRPO model still provides the correct answer \emph{Mary Had a Little Lamb}, indicating the audio is unnecessary. When the options are removed, it fails, answering \emph{Twinkle, Twinkle, Little Star} instead. This perfectly exemplifies a \textbf{Spurious} correction: the apparent improvement relies on the answer options rather than the whistled audio.

Asking the same questions in six open-ended phrasings exposes where this prior comes from. Four phrasings mention a \emph{nursery rhyme} or \emph{children's song}. The base model mostly answers \emph{Twinkle, Twinkle, Little Star}, but still varies: with real audio it once answers \emph{Baa Baa Black Sheep} and outside the options. The GRPO post-trained model answers \emph{Twinkle, Twinkle, Little Star} in all four phrasings under both real audio and silence. In the two phrasings without these terms, \emph{Twinkle, Twinkle, Little Star} never appears; both models answer outside the options. The prior is thus triggered by \emph{nursery rhyme} and \emph{children's song}, not by the audio. In other words, post-training taught the model to answer from the text, and the answer lands inside the options, which is not that we hope to learn.

\section{Conclusion}
In this work, we investigate what LALMs actually learn from different post-training methods, evaluated on audio understanding and reasoning benchmarks. Our controlled experiments reveal that reported gains stem from exploiting textual cues rather than improved audio perception: gains persist when trained on silence yet largely vanish when evaluated without options. Models simply learn a shortcut based on text priors. Because standard benchmark scores cannot distinguish this shortcut from genuine listening, our results highlight the urgent necessity for rethinking audio understanding and reasoning capabilities in current LALMs. 


\section*{Limitations}

The evidence here covers two base models, four post-training methods and six open-source models, evaluated on MMAU and MMAR. The pattern is consistent within that scope, but
it is a finite sample; other architectures, scales or benchmarks may behave differently.
We show that the performance gap between Audio Gain $\Delta\mathrm{Acc}^{a}$ and Silence Gain $\Delta\mathrm{Acc}^{s}$ stays small across independently
developed models, not that it must occur everywhere.

 

\bibliography{refs}

\appendix
\raggedbottom
\section{Global Training Configurations}
\label{app:training}

All four methods share one recipe, applied unchanged to both base models. We train LoRA
adapters of rank 32 ($\alpha=64$, no dropout) on the attention and MLP projections
(\texttt{q,k,v,o,gate,up,down\_proj}), matched by name throughout the model, so the
audio encoder's attention projections receive adapters as well as the language model's,
with AdamW at a constant learning
rate of $1\!\times\!10^{-5}$. Only the step-200
checkpoint is used in results.

For the two group-relative methods we sample $G=8$ rollouts per item at temperature 1.0
with top-$p$ 0.95, accumulate 4 groups per optimiser step, and use a KL coefficient of
0.02 against the base model with a PPO clipping range of 0.2. The micro-batch is 2
on Qwen2.5-Omni-3B and 1 on Qwen2-Audio-7B-Instruct, the only setting that differs
between the two models and one forced by memory rather than chosen.
VCRL~\cite{jiang2025vcrl} adds its variance-based group selection with $\kappa=0.75$,
warmed up from 0.3 over the first 20 steps, and an EMA coefficient of 0.9.
RFT~\cite{yuan2023rft} trains on every
verified-correct rollout of the base model at $k=8$, SFT on the gold option alone.

The reward verifies only whether the final answer matches the gold option; an unparseable
response scores zero. Training prompts carry a fixed system prompt asking the model to
listen, reason if helps, and state its answer; benchmark evaluation uses no system
prompt. For the blind conditions, \texttt{silence} zeroes each training waveform before
feature extraction, exactly as the evaluation-time silence control does, so the input preserves audio token length, and only the
information they carry is removed.

\begin{figure*}[t]
\centering
\begin{tcolorbox}[promptbox, title=\textsc{system}]
\footnotesize\ttfamily
You grade short answers to questions about an audio clip. You are given the question, the
reference answer and a candidate answer. Decide whether the candidate identifies the same
thing as the reference answer.

\vspace{5pt}
\begin{tcbraster}[raster columns=2, raster column skip=7pt, raster equal height=rows,
    colback=YesBg, colframe=YesFg, boxrule=0.6pt, arc=2pt,
    left=6pt, right=6pt, top=5pt, bottom=5pt]
\begin{tcolorbox}
\footnotesize\ttfamily
Answer yes when:\\
- the candidate is a paraphrase or synonym of the reference, or differs only in wording,
article, case, plural or spelling ("horse and buggy" for "Horse-drawn wagon");\\
- the candidate names a more specific instance of what the reference names, so that the
reference is still true of it ("vacuum cleaner" for "Mechanisms", "a woman speaking" for
"Speech");\\
- the candidate gives the reference plus extra detail that does not contradict it.
\end{tcolorbox}
\begin{tcolorbox}[colback=NoBg, colframe=NoFg]
\footnotesize\ttfamily
Answer no when:\\
- the candidate names a different thing, or a sibling of the same category rather than an
instance of it ("drums" for "Bass Guitar");\\
- the candidate contradicts the reference, is a different count, time or key, or does not
answer the question;\\
- the candidate is broader than the reference, so the reference does not follow from it
("an instrument" for "Bass Guitar").
\end{tcolorbox}
\end{tcbraster}
\vspace{5pt}

Reply with exactly one word: yes or no.
\end{tcolorbox}

\vspace{4pt}

\begin{tcolorbox}[promptbox, title=\textsc{user}]
\footnotesize\ttfamily
Question: \slot{question}\\
Reference answer: \slot{gold answer}\\
Candidate answer: \slot{model output}

\vspace{4pt}
Does the candidate answer mean the same thing as the reference answer? Answer yes or no.
\end{tcolorbox}
\caption{The judge template; the three \slot{slots} are filled per item. Blue counts as the same answer, purple does not.}
\label{fig:judgeprompt}
\end{figure*}

\section{Method-Specific Implementations}
\label{app:methods}

All four LALM post-training methods share the same underlying optimization recipe and LoRA configuration unless stated otherwise, with their specific implementations detailed below: 

\noindent\textbf{SFT.} One gradient step per batch of training items on the gold option
alone, with the same trainer, split and LoRA configuration as the other three; no
rollouts are sampled.

\noindent\textbf{RFT}~\cite{yuan2023rft}. The base model is sampled $k=8$ times per
training item at temperature 1.0, each rollout is scored by the verifier, and every
verified-correct rollout becomes a training target. Items the base model never solves
contribute nothing, so the effective pool is smaller than the full split. The rollout
pool is generated once and reused by all three seeds, so RFT's seeds differ only in data
order and initialisation.

\noindent\textbf{GRPO}~\cite{shao2024deepseekmath}. For each item a group of $G=8$
rollouts is sampled from the current policy; each rollout's advantage is its reward minus
the group mean, with no standard-deviation normalisation. Groups in which every rollout
is correct or every rollout is wrong give zero advantage and are kept rather than
filtered, so the step budget counts the same number of items for every method. Four
groups are accumulated per optimiser step.

\noindent\textbf{VCRL}~\cite{jiang2025vcrl}. Identical to our GRPO except for which
groups reach the update: a group is kept with a probability driven by its reward
variance, targeting $\kappa=0.75$, warmed up from $0.3$ over the first 20 steps with an
EMA coefficient of $0.9$. Because selection changes how many rollouts are spent per step,
we hold the step budget fixed rather than the rollout budget.

\noindent\textbf{Blind training.} The blind conditions change nothing above.
\texttt{silence} zeroes each waveform before feature extraction, so the input keeps its
audio tokens and its length; \texttt{none} drops the audio turn from the chat template
entirely, which also shortens the prompt. This paper's blind results use
\texttt{silence}.

\section{Details of Open-Ended Evaluation}
\label{app:openend}

To evaluate models under open-ended settings without changing the underlying intent, we systematically adapt multiple-choice questions and employ a standardized LLM-based judge, as detailed below: 

\noindent\textbf{Rewriting the questions.}
Stripped of its options, a question that points at them becomes unanswerable for a reason
that has nothing to do with audio: \emph{identify which of the following sounds can be
heard for the longest duration} names a list that is no longer there. The rewriting is
purely syntactic. A trailing instruction to pick from the list is deleted
(\emph{\ldots{} Choose the correct option from the following options:}); \emph{which of
the following} contracts to \emph{which} before a noun and to \emph{what} before a verb,
so \emph{identify which of the following is the dominant sound} becomes \emph{identify
what is the dominant sound}; a remaining reference to the list or to the options above is
deleted. A phrase that points at the stimulus rather than at an answer set
(\emph{the following audio}, \emph{these two recordings}) is exempt. The rules touch 172
of the 1{,}000 MMAU questions and 407 of MMAR's, and never change what is asked.

Six questions survive the rewriting still presupposing an option list, four in MMAU and
two in MMAR. 
Some are negations (\emph{which genre is NOT part of the audio's style} has no answer without a list to exclude from) and the rest are items the rules flag conservatively, such as a question about which option a speaker in the audio wants chosen. 
We ask and
score them unchanged rather than drop them, so that every protocol in this paper runs
over the same 1{,}000 items per benchmark. They cost almost nothing either way: base and
post-trained models miss them alike, and removing them from the denominator moves no
checkpoint's open-ended accuracy by more than $0.13$ points on MMAU or $0.07$ on MMAR
(median $0.05$ over the 132 judged runs), leaving $\Delta\mathrm{Acc}^{a}$ and $\Delta\mathrm{Acc}^{s}$ unchanged.

\noindent\textbf{The judge.}
One model, Qwen3.6-27B, grades every response in this paper, decoding greedily for at
most eight tokens with its reasoning mode off. It never receives the audio, so it is scoring
agreement between two strings and not listening on the model's behalf.
Figure~\ref{fig:judgeprompt} gives the template in full: a fixed system message that
defines when two answers count as the same, and a user message whose three slots are
filled per item. A verdict containing neither word is scored as no; across the
132{,}000 judgements in this paper there were none.

\begin{figure}[t]
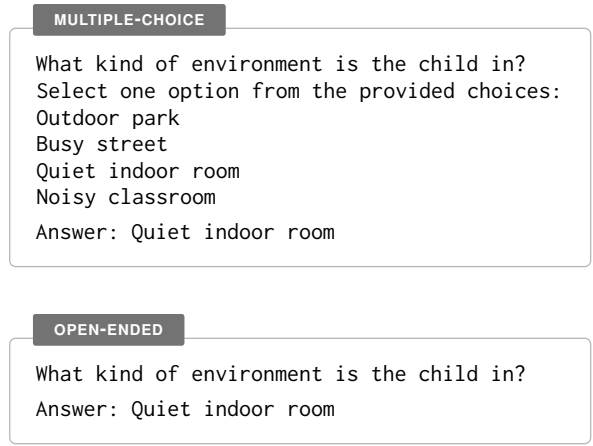

\centering
\begin{tcolorbox}[promptbox, title=\textsc{multiple-choice}]
\footnotesize\ttfamily
What kind of environment is the child in?\\
Select one option from the provided choices:\\
Outdoor park\\
Busy street\\
Quiet indoor room\\
Noisy classroom\\[3pt]
Answer: Quiet indoor room
\end{tcolorbox}

\vspace{4pt}

\begin{tcolorbox}[promptbox, title=\textsc{open-ended}]
\footnotesize\ttfamily
What kind of environment is the child in?\\[3pt]
Answer: Quiet indoor room
\end{tcolorbox}
\caption{One training item in both forms.}
\label{fig:trainprompt}
\end{figure}

\section{Details of Open-Ended Post-Training}
\label{app:opentrain}

Training models in open-ended settings requires adapting the question pool and redefining the reward function, implemented as follows:

\noindent\textbf{The question.}
This pool renders its choices inside the question text, after a fixed header, and leaves
the \texttt{options} field empty, so the list is cut out of the string rather than dropped
from a field. What remains is rewritten by the rules of Appendix~\ref{app:openend}, the
same ones the open-ended evaluation uses. The header is present in all $12{,}267$ items
and nothing of the option text survives the cut. Figure~\ref{fig:trainprompt} shows an
item in both forms.

\noindent\textbf{The reward.}
The multiple-choice reward verifies one thing: whether the final answer matches the gold
option. 
Without options there is no option to match, so the reward has to check the gold \emph{answer} instead, and the obvious version of that is far too strict to train on. 
\emph{The child is in a quiet indoor room} is a correct
answer to its question and scores zero against the gold string \emph{Quiet indoor room},
and that matters more than it would for an evaluation metric: GRPO's advantage is a reward
minus its group mean, so a group of eight rollouts that all score zero contributes no
gradient at all.

The reward therefore accepts a response that \emph{contains} the gold answer. 
It first tries equality after normalisation, which is the multiple-choice rule unchanged, and then looks for the gold string inside the response. 
The search is on word boundaries rather than by substring, which is the same
rule Section \ref{sec:setting}'s verifier already uses to recover an option from free text:
\emph{male} is a substring of \emph{female}, and plain containment would make every answer
about a female speaker ambiguous. Everything else is as in Appendix~\ref{app:training}:
the same audio, the same schedule, the same two hundred steps. Under this reward
the mean settles between $0.3$ and $0.5$ over training, against about $0.5$ for the
multiple-choice reward on the same pool.

\end{document}